\documentclass{article}

\usepackage{microtype}
\usepackage{graphicx}
\usepackage{subfigure}
\usepackage{booktabs} 
\usepackage{amsmath}

\usepackage[normalem]{ulem}
\usepackage{multirow}
\usepackage[colorinlistoftodos,prependcaption,textsize=tiny]{todonotes}

\newcommand{\rc}[1]{Rawcopy}

\newcommand{\vgg}[1]{BL VGG}
\newcommand{\densenet}[1]{BL DenseNet}
\newcommand{\dildensenet}[1]{BL Dilated DenseNet}
\newcommand{\lstmdensenet}[1]{BL LSTM DenseNet}
\newcommand{\deepsnp}[1]{Deep SNP}

\newcommand{\evgg}[1]{\emph{BL VGG}}
\newcommand{\edensenet}[1]{\emph{BL DenseNet}}
\newcommand{\edildensenet}[1]{\emph{BL Dilated DenseNet}}
\newcommand{\elstmdensenet}[1]{\emph{BL LSTM DenseNet}}
\newcommand{\edeepsnp}[1]{\emph{Deep SNP}}

\usepackage[accepted]{icml2018}

\icmltitlerunning{Deep SNP: An End-to-end Deep Neural Network for Break-point Detection in SNP Array Genomic Data}

\begin{document}

\twocolumn[
\icmltitle{Deep SNP: An End-to-end Deep Neural Network with Attention-based Localization for Break-point Detection in SNP Array Genomic data}
\icmlsetsymbol{equal}{*}

\begin{icmlauthorlist}
\icmlauthor{Hamid Eghbal-zadeh}{equal,cp}
\icmlauthor{Lukas Fischer}{equal,scch}
\icmlauthor{Niko Popitsch}{anna}
\icmlauthor{Florian Kromp}{anna}
\icmlauthor{Sabine Taschner-Mandl}{anna}
\icmlauthor{Khaled Koutini}{cp}
\icmlauthor{Teresa Gerber}{anna}
\icmlauthor{Eva Bozsaky}{anna}
\icmlauthor{Peter F. Ambros}{anna}
\icmlauthor{Inge M. Ambros}{anna}
\icmlauthor{Gerhard Widmer}{cp}
\icmlauthor{Bernhard A. Moser}{scch}
\end{icmlauthorlist}

\icmlaffiliation{cp}{Institute of Computational Perception,
Johannes Kepler University, Linz, Austria}
\icmlaffiliation{scch}{Software Competence Center Hagenberg (SCCH), Hagenberg, Austria}
\icmlaffiliation{anna}{Children's Cancer Research Institute (CCRI), Vienna, Austria}
\icmlcorrespondingauthor{Hamid Eghbal-zadeh}{hamid.eghbal-zadeh@jku.at}
\icmlkeywords{Break-point detection, SNPArray, Convolutional Neural Networks, Attention mechanism, Weak labels}

\vskip 0.3in
]
\printAffiliationsAndNotice{\icmlEqualContribution} 

\begin{abstract}
Diagnosis and risk stratification of cancer and many other diseases require the detection of genomic breakpoints 
as a prerequisite of calling copy number alterations (CNA). This, however, is still challenging and requires 
time-consuming manual curation. As deep-learning methods outperformed classical state-of-the-art algorithms in various
domains and have also been successfully applied to life science problems including medicine
and biology, we here propose Deep SNP, a novel Deep Neural Network to learn from genomic data.
Specifically, we used a manually curated dataset from 12 genomic single nucleotide polymorphism
array (SNPa) profiles as truth-set and aimed at predicting the presence or absence of genomic breakpoints, 
an indicator of structural chromosomal variations, in windows of 40,000 probes.
We compare our results with well-known neural network models as well as \rc , 
though this tool is designed to predict breakpoints and in addition genomic segments with high sensitivity. 
We show, that Deep SNP is capable of successfully predicting the presence or absence of a breakpoint 
in large genomic windows and outperforms state-of-the-art neural network models. 
Qualitative examples suggest that integration of a localization unit may enable breakpoint detection and prediction of genomic segments, even if the breakpoint coordinates were not provided for network training. These results warrant further evaluation of DeepSNP for breakpoint localization and subsequent calling of genomic segments.

\end{abstract}

\section{Introduction}
\label{sec:intro}

Copy-number alterations (CNA) such as losses, gains or amplifications of DNA sequences result in changes in the copy number status (CNS) of the respective genomic regions. If CNAs occur within chromosomes, this results in segmental CNAs and they can range in size from a few base pairs to whole chromosome arms and are, for example, used to molecularly diagnose and/or stratify cancer patients into risk-groups, a pre-requisite to allocate patients to appropriate treatment protocols.
Accurate identification of CNAs is thus critical for studying the pathogenesis of cancer and many other diseases.

Today, despite emerging new technologies such as whole-genome sequencing (WGS), microarrays and in particular single nucleotide polymorphism array (SNPa) are a common choice for copy number analyses in clinical routine. Reasons for this arguably include their simplicity, robustness and good  cost-benefit ratio. SNPa interrogate a patient genome at defined genomic positions using copy number and allele-specific oligonucleotide probes.

A core step of copy-number analyses is the segmentation of patient genomes into segments of equal copy number. Such genomic segments are defined by their endpoints (aka breakpoints), i.e. transitions in the copy number state (CNS) of two adjacent genomic segments, and can be determined by considering neighboring probes of a SNPa. These probes provide information about the relative abundance of DNA at a particular genomic position (i.e., its copy-number) and about the directly related ratio of two different alleles at polymorphic (SNP) positions (the B-allele frequency) \cite{LaFramboise_SNPaReview_2009}.

Accurate copy number calling from these data requires breakpoint detection, but is challenging for multiple reasons including mosaicism, technical noise, repetitiveness of the genome, and technology intrinsic biases such as guanine-cytosine (GC) 
bias (DNA fragments that are very rich/poor in G and C bases show different hybridization/amplification characteristics, cf. \cite{Benjamini_GCbias_2012}) or probe cross-hybridization.  

In this paper we propose a novel end-to-end deep neural network namely Deep SNP, specialized to process raw SNP array genomic data and to predict breakpoints therein.
Deep SNP is capable of processing very long stretches of copy number data by incorporating dilated convolution layers and it benefits from state-of-the-art feature learning architectures. In addition, it can learn long genomic distance relations in its distance embedding space using recurrent layers. Finally, by integrating attention-based localization layers, Deep SNP is capable of accurately pinpointing breakpoints without the need of using accurate labels for training and reduce the recall and increase breakpoint calling precision within the given genomic windows of 40,000 probes, compared to the general purpose state-of-the-art deep models. Though direct comparison is difficult, Deep SNP also compares well to biological tools such as \rc~ \cite{Mayrhofer2016} in predicting the presence of breakpoints.

\section{Related Work}
\label{sec:related_work}

Over time, many different algorithms for breakpoint detection have been proposed, mostly relying on statistical methods including Hidden Markov Models (HMM) \cite{Wang_PennCNV_2007,Marioni_BioHMM_2006}, Bayesian approaches \cite{Pique-Regi_BayesCNV_2008,Zhang_BayesCNV_2010} or circular binary segmentation (CBS; \cite{Olshen_CBS_2004}), a simple recursive method that was shown to outperform other approaches in terms of sensitivity and false discovery rate (FDR; \cite{Zhao_CNVReview_2013}).
Briefly, CBS starts by considering a whole chromosome and recursively segments it by applying a simple statistical test (maximal t-statistic) for change point detection to these segments. CBS stops when no more change points can be found in the current segmentation \cite{Venkatraman_CBS2_2007}. 
HMM and CBS based solutions are implemented in many popular copy number variation (CNV) tools such as ChAS, Nexus or \rc{} and provide relatively accurate and stable copy-number calls.

Current algorithms perform well in routine analysis, when DNA quality and quantity are sufficient and tumor cell content is high and the respective SNPa data shows relatively ,,clean'' profiles. 
In practice, however, SNPa profiles are often noisy and reasons for this include
cross-linked and/or fragmented DNA due to 
tissue preparation procedures such as formalin fixation or unusually long storage of tissue/DNA. Furthermore, when DNA amounts are limited, such as in cell-free DNA extractions from liquid biopsies, low DNA input and contamination with non-tumor DNA can also contribute to the noise in the data. 
With current algorithms, noisy and/or subclonal/contaminated samples result often in reduced segmentation accuracy, mainly due to increased false positive rates, and/or high fragmentation (i.e. many small neighboring segments of alternating copy number). Such segments require manual curation which is work intensive and error prone. Thus, there is a need to develop algorithms predicting breakpoints and genomic segments with higher precision from these data.

Deep learning methods outperformed classical state-of-the-art algorithms across various domains including image classification, speech recognition, language translation or document analysis and were recently also successfully applied to various problems in the life science domain \cite{Angermueller2016}. Deep learning architectures can learn unknown or hidden relationships from highly complex and noisy data if a sufficient amount of training data is available, even if only weak labels can be provided~\cite{Ching2018}. 
DeepVariant, a deep learning architecture to predict SNVs or small insertions/deletions (indels) in WGS data was proposed by~\cite{Poplin2016}. Although results show that DeepVariant can learn the statistical relationship between aligned reads and true SNVs/indels over various sequencing technologies, the architecture is not directly transferable to be applied to call CNVs from SNPa data as technology and thus data structure and resolution differ substantially. Moreover, the architecture is a pure classification approach relying on currently available frameworks calling candidate predictions. Thus, DeepVariant is not suitable to further call CNVs. 
Most recently, ~\cite{gupta_dilated_2017} proposed to use dilated convolution on regulatory marker locations from ENCODE~\cite{consortium_integrated_2012} to model long-distance genomic dependencies. They showed that dilated convolution can outperform LSTM~\cite{hochreiter_long_1997} recurrent models. Nevertheless they did not evaluate their method on detection of rare genomic events such as breakpoints. In addition, their method requires long input sequences and does not have a localization ability.

\section{Data Collection and Preprocessing}
\label{sec:data}

For our initial tests we analyzed 12 samples of 5 neuroblastoma patients (material was either tumor, or disseminated tumor cells (DTCs)) on the Affymetrix CytoScan HD SNP array platform \cite{Ambros2014UltraHighDS}. The resulting CEL files contain $\sim2.8$ Mio raw array intensity values which were converted into normalized log ratio (LRR) and B-allele frequency values (BAF) using \rc{}, an open R package for processing Affymetrix microarray data \cite{Mayrhofer2016}. 
\rc{} first calculates raw LRR and BAF values according to the following formulas where $\overline{A}$ and $\overline{B}$ are mean intensities of the respective SNPa probes (see \cite{Mayrhofer2016} for details):
\begin{equation}
\label{eq:logr}
	LRR = log_2 \sqrt{\overline{A^2} + \overline{B^2}}
\end{equation}
\begin{equation}
\label{eq:raw_baf}
	BAF = \frac{\overline{B}}{\overline{A} + \overline{B}}
\end{equation}
These two measures are then further normalized using observed per-probe value distributions derived from a large set of reference samples. LRR values are furthermore corrected for fragment length and GC content. 
Finally, \rc{} calls breakpoints using an allele-specific CBS algorithm and outputs text files containing normalized LRR and BAF values as well as called copy number segments. These resulting data files contained $2,819,443$ LRR values and around $480,000$ BAF values each (Figure \ref{fig:data_gen}, (a)). 

To create the ground truth, we manually curated the \rc{} predictions using our in-house editor Varan-GIE\footnote{https://github.com/popitsch/varan-gie}, taking additional data files (such as related datasets, mappability tracks and annotations of common CNVs) into account. Highly fragmented segments often occur in noisy profiles as artifacts predicted by \rc{} and need to be merged. In case of false negative segment predictions, additional segments have to be generated and added to the final segmentation. More frequently, due to a high sensitivity but lower specificity of \rc{} predictions, false positive segments have to be removed. The manually curated segments were called by experienced biologists and always double-checked by at least one additional person (according to the lab routine) and considered as the truth-set in the following.
To train our network, we combined the \rc{} output, our manual curation set and additional genome-wide data to create genome-wide data files that contain the following data:
1) Genomic position (chromosome + offset) of the respective SNPa probe, 
2) \rc{} LRR value, 
3) \rc{} BAF (-1 encoding missing values), 
4) Encoded truth-set copy number state (normal, loss, gain, amplification) at the probe position, 
and 5) Encoded \rc{} copy number state at the probe position (used in our evaluation).
\subsection{Data Preparation for Deep Learning}
\label{subsec:data_preparation_dl}
In this section, we explain the details of data preparation we have done to feed the SNPa data to our deep neural networks.
The samples of SNPa are very long ($\sim2.8$ Mio. values) and can not be directly fed to a neural network to process. Therefore, we designed a windowing scheme to create training and validation data for the neural network systems.
Using this procedure, based on the truth-set of copy number state we select positive and negative windows of probes and used them as training and validation data.
A window of $n$ probes is considered a positive window, if within that window the copy number state changes at least once.
Also a window of $n$ probes is considered a negative window, if the copy number state does not change at all within that window.
The change of copy number states (copy number state transition) represents a breakpoint, therefore from now on, a positive window will be referred to as a window with at least one breakpoint.

For each genomic sample, we select the positive windows as follows: First we locate the position of all breakpoints in that sample based on the truth-set. Second, a window of $n$ probes is centered at all the known breakpoint positions one by one. 
Due to the possible small distance between neighboring breakpoints, positive windows can contain multiple breakpoints. 

To select negative windows, we randomly select (non-overlapping) windows with $n$ consecutive probes that do not contain any breakpoints. LRR and BAF values were concatenated in each window and added to create a $2 \times n$ feature vector representing one window. 

\begin{figure}[!ht]
\vskip 0.2in
\begin{center}
\centerline{\includegraphics[width=\columnwidth]{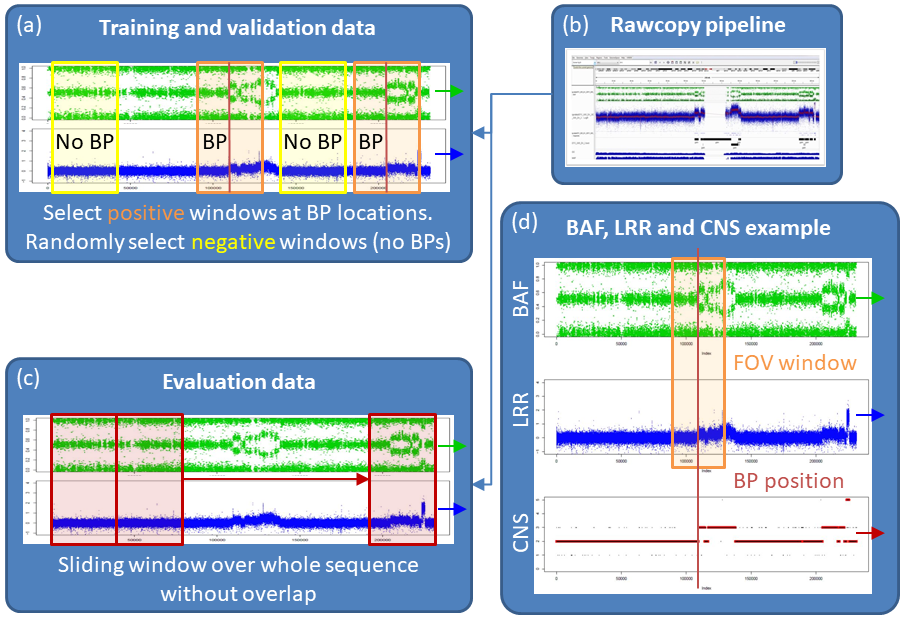}}
\caption{Data generation pipeline. All images show only a small part of the whole sample sequence with $\sim225,000$ values. (a) Positive data for training and validation data is created by placing windows at breakpoint (BP) positions. Negative data is created by placing windows at random positions where no BPs are located (e.g. the beginning and/or end of the sequence as well as in between breakpoints. BAF, LRR and truth-set CNS are then extracted in each defined window and concatenated to form the respective input data. (b) Example UI (IGV/VARAN-GIE) of the \rc{} pipeline, which provides BAF, LRR values and \rc{} CNS, and enables manual curation. (c) Evaluation data is generated by sliding a window over a full sample sequence ($\sim2.8$ Mio values) without overlap (figure shows a subset). Again BAF and LRR values are extracted in each defined window and concatenated. (d) Example window (orange rectangle) placed at a BP position (red line). CNS in the last row depicts the manual truth-set CNS (red horizontal lines) as well as the \rc{} CNS (black horizontal lines and dots).}
\label{fig:data_gen}
\end{center}
\vskip -0.2in
\end{figure}

No further information (except breakpoint coordinates extracted at positions of copy number state transition) such as chromosome position was taken into account for window selection. 
Finally, each feature vector was labeled as "has breakpoint(s) / positive", if at least one breakpoint is located within the window or "has no breakpoint(s) / negative", otherwise. The training and validation data selection is depicted in Figure \ref{fig:data_gen}, (a).
The evaluation data set is created by sliding a window with no overlap over the whole sample sequence as shown in Figure \ref{fig:data_gen}, (c) (for better presentability only a small subset is shown).

For training, validation and evaluation we chose a window size of $n=40,000$ probes. (see Section \ref{subsec:evaluation}).

\subsection{Cross-Validation Folds}
\label{subsec:cross_validation}
In order to evaluate each model on all available samples, we created a 4-fold cross validation. We divide our 12 samples into 4 separated folds such that each fold consists of 3 samples. In each round of training, the models are trained on 9 samples and tested on the remaining 3.

\section{Deep SNP: An End-to-end Neural Network for Breakpoint Detection in SNP Array}
\label{sec:genome_net}

The genomic data considered by our approach is characterized by very long sequences of low-dimensionality data which makes analysis with conventional deep learning architectures difficult.
In our empirical results we show that state-of-the-art architectures such as VGG~\cite{simonyan2014very} and DenseNet~\cite{huang_densely_2016} that perform very well in audio~\cite{eghbal2016cp,hershey2017cnn} and image processing~\cite{simonyan2014very,huang_densely_2016} applications are not capable of coping well with these data.
Hence, we designed an architecture with specialized units targeting specific processing goals to deal with these challenges.
In the following section, we detail the architecture and different units used in Deep SNP.

\subsection{Network Architecture}

Deep SNP is a deep neural network trained with stochastic gradient descent (SGD) in an end-to-end fashion. Deep SNP has 5 main units and each unit plays an important role for learning a suitable hidden representation from the genomic data and further learn the relevant aspects of a breakpoint from the long, low-dimensional input sequence.

\begin{figure}[ht]
\vskip 0.2in
\begin{center}
\centerline{\includegraphics[width=\columnwidth]{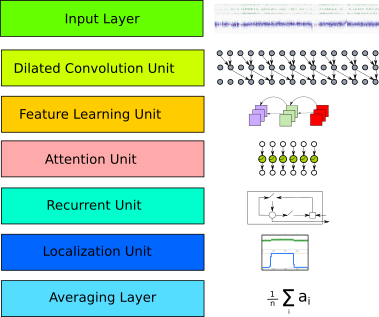}}
\caption{Block-diagram of the proposed Deep SNP.}
\label{fig:diagram}
\end{center}
\vskip -0.2in
\end{figure}

The Deep SNP units as shown in Figure~\ref{fig:diagram} are as follows: 
1) Dilated Convolutional Unit: to cope with the very long SNPa sequences and learn from very low-dimensional data.
2) Feature Learning Unit: To learn a reasonable hidden representation suitable for breakpoint detection.
3) Attention Unit: To be able to focus on specific regions, or use a wide-range of probes.
4) Distributed Recurrent Unit: To learn the dependencies in the learned hidden representation, between hidden factors of different genomic positions, and finally 
5) Localization Unit: To be able to decide which of the hidden factors of the genomic positions are more relevant to the task.

\paragraph{Dilated Convolutional Unit} Dilated convolutional layers~\cite{yu2015multi} are specialized convolutional layers  with large receptive fields that can process very long inputs.
These layers are known for their significant performance in audio processing and machine translation using long sequences of one-dimensional data, to learn from or generate raw audio/textual input instead of 2D image-like features such as spectrograms. 
Good examples for successful applications of dilated convolutions in dealing with very long one-dimensional sequences are ByteNet~\cite{kalchbrenner2016neural} for fast machine translation, and Wavenet~\cite{van2016wavenet} which is the state-of-the-art in speech synthesis and is capable of synthesizing directly very long audio samples, while it is only trained on long sequences of raw audio. Recently,~\cite{gupta_dilated_2017} their were used to process Genomic data and they have shown great promise.

\paragraph{Feature Learning Unit} In this unit, we incorporate and adapt state-of-the-art convolutional architectures which can efficiently learn a high-level representation. We decided to choose a state-of-the-art architecture for image recognition called DenseNet~\cite{huang_densely_2016}. Nevertheless, this unit is interchangeable with other architectures such as ResNet~\cite{he2016deep} or VGG~\cite{simonyan2014very}.
We connect the input layer of the feature learning unit to the representation that our dilated convolutions learned.
We modified the DenseNet in order to learn from the genomic position dimension which is very long. Therefore, we only used 1D filters and 1D pooling layers.

\paragraph{Attention Unit} Biologists often have to spend long hours investigating the probes carefully to be able to annotate a breakpoint as the truth-set. They need to use special tools to explore such long sequences, enabling them to \emph{zoom-in} and \emph{zoom-out} in specific regions of the genomic data to make the final decision about a breakpoint.
We use an \emph{Attention Unit} that allows our network to attend on any activations in any desired probe index. This enables Deep SNP to explore the probes and focus on specific regions of the genomic sequence if necessary and process the data in similar ways as the biologist annotators.

\paragraph{Distributed Recurrent Unit} Recurrent models such as LSTM~\cite{hochreiter_long_1997} and GRUs~\cite{chung2014empirical} are well-known models to model sequential data such as audio, text and genome sequences~\cite{gupta_dilated_2017}. 
Therefore, we use a bidirectional Gated Recurrent layer (GRU) and apply it  on the hidden space that our feature learning module learned. We prefer GRU over LSTM because of its simplicity and competitive performance over LSTM.
We apply the GRUs on the sequence of hidden activations formed by the dimension in the hidden space that represents the genomic position of the features.
As the filters used are 1D, the dimension in the hidden representation that represents the genomic position can be tracked back to the input and represent its corresponding input probes as it is explained in the localization unit.

\paragraph{Localization Unit} Localization Units can provide a mechanism to trace a hidden representation back to the input in a specific dimension. By using distributed units we can repeat a processing unit such as softmax on every node of a specific dimension. This way, the predictions of each node in that dimension represent the corresponding indexes in the input (which directly maps to the respective genomic position).

We tried three different localization mechanisms for localizing the regions in the input probes with high importance for breakpoint prediction. 
The first, is to apply a distributed dense layer with softmax activations on the dimension of the hidden representation learned from the previous unit on the dimension that represents the genomic position. Then each softmax output can be interpreted as the importance of a series of consecutive probes in the input. For the final prediction, these softmax probabilities are averaged into a final probability and can be used to train the model as there is only one label provided per window.
Nevertheless, the output of the distributed softmax before averaging can be used for localization purposes. These results can be found under \emph{No Attention} in Table~\ref{tab:empirical_results}.

For our second localization method, we use a technique from state-of-the-art in audio event detection with weakly-labeled data~\cite{xu2017large}.
We added an additional attention gate using a distributed dense layer with sigmoid activations that can control the output of the distributed softmax probability by multiplying the output of the sigmoid gate to the softmax prediction. This mechanism allows the network to close the activations related to the genomic position that are not beneficial for the breakpoint predictions.
Similar to our first localization unit, the outputs are averaged and used for training. The output of the attention gate can be then used for localization purposes separately.  These results can be found under \emph{Final Attention} in Table~\ref{tab:empirical_results}.

For our last localization, we additionally added the attention gate to the input of the recurrent unit. This helped to improve the localization performance, as the inputs can be controlled even before processed by the GRUs.
These results can be found under \emph{Mid+Final Attention} in Table~\ref{tab:empirical_results}.
Examples of the localization unit using Mid+Final attention are provided in Figure~\ref{fig:localisation_preds}.

\section{Results}
\label{sec:results}
In this section, we provide the empirical results in Table~\ref{tab:empirical_results} for SNPa breakpoint classification using different methods. In addition, we use Deep SNP's localization unit to pinpoint the position of the breakpoint in a window and illustrate breakpoint and genomic segment positions with example views. For these results, \emph{the same} Deep SNP model as used in Table~\ref{tab:empirical_results} is used which was only trained on binary labels per window and basically had no knowledge about the exact position of the breakpoint(s) within the window.
These results are shown in Figure~\ref{fig:localisation_preds}.

In Section~\ref{subsec:baseline} and Section~\ref{subsec:genet_training}, we explain the baseline systems and clarify the training procedure in the baselines as well as the proposed method.

\subsection{Baselines}
\label{subsec:baseline}
This section explains the other network architectures we evaluated on, known as \emph{Baseline methods}.
We used two state-of-the-art feed forward deep neural networks namely VGG~\cite{simonyan2014very}, and DenseNet~\cite{huang_densely_2016}) as baseline methods to demonstrate the learning abilities of \deepsnp{} compared to general purpose neural networks.
We also compared our results to the breakpoints detected by \rc{}, a specialized tool for CNV detection.
 
\subsubsection{\rc{}}
\label{subsubsec:raw_copy_baseline}
\rc{} data processing was done according to Section \ref{sec:data}. The encoded copy numbers at the probe positions were then evaluated against the manually curated truth-set.
\rc{} uses various built-in reference data precompiled from a large number (2875 samples, Supplementary Table 1~\cite{Mayrhofer2016}) of ethnically diverse samples, with variations also in technical quality to improve LRR and BAF normalization.
Therefore, in the sense of training data, \rc{} benefited from significantly more data compared to Deep SNP and the rest of the Deep Learning baselines that were only trained with 9 out of 12 available samples.

To evaluate \rc{} for breakpoint prediction, first the copy number states were predicted for each probe and further converted to our binary format as explained in Section~\ref{subsec:data_preparation_dl}.

\subsubsection{Deep Neural Networks}
\label{subsec:dnns}
\paragraph{Feed-forward Neural Networks}
We used three feed-forward deep neural network architectures that are known to perform reasonable well in various tasks such as image recognition~\cite{huang_densely_2016} and audio acoustic scene classification~\cite{eghbal2016cp,hershey2017cnn} as baselines.
We use two well-known  deep convolutional feed forward architectures namely VGG~\cite{simonyan2014very} used in \evgg{} and DenseNet~\cite{huang_densely_2016} used in \edensenet{}.
We adapted the filter sizes of the aforementioned baselines to the task and the data used. 
Also no dilation, attention, recurrent units or localization units were used in these baselines.

\edildensenet{} is a baseline designed based on the proposed method in~\cite{gupta_dilated_2017} which uses incremental dilation in a feed-forward CNN to model long-distance genomic dependencies.
We adapt their architecture to our task (as their task was not detection) by adding convolutional layers followed by a global average-pooling layer to play the role of a "classifier" at the final part of the network.
We further improve the CNN architecture by upgrading it to a DenseNet architecture with incremental dilation.
Further, this design is also aligned with other DenseNet architectures in \edensenet{} and the proposed Deep SNP in terms of filter sizes, 
number of parameters and the overall architecture design.
These changes resulted in significant improvements in our task, compared to the initial architecture.
The architecture of the implemented \evgg{}, \edensenet{} and \edildensenet{} are shown in the appendix (Table~\ref{tab:baselines_architectures}).

\paragraph{Recurrent Neural Networks}
Recurrent Neural Networks (RNNs) are known for their ability to model sequential data. Therefore, we use a convolutional bidirectional LSTM~\cite{hochreiter_long_1997} as our final baseline.
We design this baseline based on the results reported in~\cite{gupta_dilated_2017}, as authors replaced incremental dilated convolutions with convolutional layers
followed by bi-directional LSTM after the convolutional layers.
Similarly, we design our last baseline \elstmdensenet{} based on a model used~\cite{gupta_dilated_2017}.
We replace the convolutional layers with a DenseNet architecture and we add a bidirectional LSTM for sequence modeling of the learned representation by the convolutional layers.
No dilation, attention, or localization units were used in this baseline.
We then add the "classifier" part of the network similar to the other baselines using a convolutional layer followed by global average-pooling.
The architecture of the \elstmdensenet{} is shown in Table~\ref{tab:baselines_architectures}.

All baselines were trained for 200 epochs with a batch size of 25 using AMSGrad~\cite{reddi_convergence_2018} (an adaptive version of SGD), categorical cross-entropy loss, initial learning rate of 0.001 which was reduced on plateau by 0.1.
The patience of 50 and early stopping of 100 epochs was carried out as well.
Deep SNP as well as deep learning baselines are implemented in Keras~\cite{chollet2015keras} and the training is done on a Nvidia DGX Station using Keras with Tensorflow~\cite{abadi2016tensorflow} backend.

\subsection{Deep SNP Training Strategy}
\label{subsec:genet_training}
The proposed Deep SNP was trained for 100 epochs with the same hyper-parameters, optimizer and training strategy as the baseline networks (see Section \ref{subsec:dnns}). The architecture of Deep SNP can be found in Table~\ref{tab:genet_architecture}.

\subsection{Evaluation}
\label{subsec:evaluation}
All described algorithms were evaluated using 4-fold cross-validation. For Deep SNP and deep learning baselines, in each fold nine samples were used for training and three for validation (selected windows) and evaluation (full sequence). 
Note that evaluation data was created differently than validation data in terms of the length of data used (see Section \ref{sec:data}). 
For each fold we report F1-Score (F1), Precision (PREC) and Recall (REC). For F1, PREC and REC we apply averaging in two different ways: macro (MAC - calculate metrics for each label, and find their unweighted mean) and binary (POS - calculate metrics only for positive labels i.e. with breakpoint).
The window size for training sample selection was $40,000$. This value was chosen based on empirical experiments conducted during algorithm development.

\subsection{Empirical Results}
\label{subsec:empirical_results}
The results for each evaluated algorithm are shown in Table~\ref{tab:empirical_results}. 
\begin{table}[!htbp]
\caption{Cross-validation results for all conducted experiments. All values are shown in \%. Binary precision and recall (PREC POS, REC POS) are highlighted in all experiments.}
\label{tab:empirical_results}
\vskip 0.15in
\begin{center}
\begin{small}
\begin{sc}
\begin{tabular}{l|ccccr}
\toprule
(\%) / Folds& Fold 1 & Fold 2 & Fold 3 & Fold 4 \\
\midrule
\multicolumn{5}{c}{\rc{}}\\
\hline
F1 Mac	&43.42	&48.30	&23.65	&41.49\\
F1 Pos	&17.39	&46.60	&39.06	&37.37\\
Prec Mac&55.40	&48.36	&26.76	&41.78\\
Prec Pos&\underline{09.71}	&\underline{30.57}	&\underline{24.27}	&\underline{23.12}\\
Rec Mac &68.53	&65.75	&52.15	&63.54\\
Rec Pos &\underline{83.33}	&\underline{97.96}	&\underline{100.00} &\underline{97.37}\\
\hline
\hline
\multicolumn{5}{c}{BL VGG}\\
\hline
F1 Mac	 	&51.75	&	22.51	&	42.12	&	45.10	\\
F1 Pos 		&11.43	&	28.70	&	0.0	&	0.0	\\
Prec Mac 	&85.45	&	23.0	&	72.77	&	82.16	\\
Prec Pos 	&\underline{8.70}	&	\underline{18.23}	&	\underline{0.0}	&	\underline{0.0}	\\
Rec Mac  	&53.11	&	38.55	&	47.55	&	50.0	\\
Rec Pos  	&\underline{16.67}	&	\underline{67.35}	&	\underline{0.0}	&	\underline{0.0}	\\
\hline
\multicolumn{5}{c}{BL DenseNet}\\
\hline
F1 Mac	 	&54.54	&	43.05	&	43.20	&	51.84	\\
F1 Pos 		&12.50	&	0.0	&	0.0	&	13.64	\\
Prec Mac 	&93.43	&	75.59	&	76.06	&	82.16	\\
Prec Pos 	&\underline{25.0}	&	\underline{0.0}	&	\underline{0.0}	&	\underline{50.00}	\\
Rec Mac  	&53.42	&	49.09	&	49.69	&	53.09	\\
Rec Pos  	&\underline{8.33}	&	\underline{0.0}	&	\underline{0.0}	&	\underline{7.89}	\\
\hline
\multicolumn{5}{c}{BL Dilated DenseNet}\\
\hline
F1 Mac	 	&70.34	&	84.29	&	58.47	&	76.72	\\
F1 Pos 		&44.44	&	74.70	&	29.03	&	60.32	\\
Prec Mac 	&92.96	&	90.14	&	79.34	&	88.26	\\
Prec Pos 	&\underline{40.0}	&	\underline{91.18}	&	\underline{75.0}	&	\underline{76.0}	\\
Rec Mac  	&72.76	&	80.72	&	58.08	&	73.29	\\
Rec Pos  	&\underline{50.0}	&	\underline{63.27}	&	\underline{18.0}	&	\underline{50.0}	\\
\hline
\multicolumn{5}{c}{BL LSTM DenseNet}\\
\hline
F1 Mac	 	&73.77	&	79.60	&	77.68	&	73.27	\\
F1 Pos 		&50.0	&	66.67	&	63.01	&	53.57	\\
Prec Mac 	&95.31	&	87.79	&	87.32	&	87.79	\\
Prec Pos 	&\underline{62.50}	&	\underline{89.66}	&	\textbf{\underline{100.0}}	&	\underline{83.33}	\\
Rec Mac  	&70.09	&	75.62	&	73.0	&	68.88	\\
Rec Pos  	&\underline{41.67}	&	\underline{53.06}	&	\textbf{\underline{46.0}}	&	\underline{39.47}	\\
\hline
\hline
\multicolumn{5}{c}{Deep SNP (No attention)}\\
\hline
F1 Mac&  90.86	&87.27&	65.50	&86.23\\
F1 Pos&  84.21	&80.0	 &  34.78	&80.0\\
Prec Mac&95.71	&91.43&	92.86	&89.05\\
Prec Pos&\underline{96.0}	    &\underline{75.0}	 &  \underline{80.0	}&\underline{79.31}\\
Rec Mac & 87.22	&89.28&	60.85	&86.42\\
Rec Pos & \underline{75.0} &\underline{85.71}& \underline{22.22}	&\underline{80.70}\\
\hline
\multicolumn{5}{c}{Deep SNP (Final attention)}\\
\hline
F1 Mac	 	& 88.84 &	 93.36	 &69.39	 &	88.50 \\
F1 Pos 		& 80.70	 &	 89.41	 &42.86   &	83.18\\
Prec Mac 	& 94.77 &	 95.71	 &92.38	 &	90.95 \\
Prec Pos 	& \underline{92.0} &		 \underline{88.37}	 &\underline{60.0}  &	\underline{83.93 }\\
Rec Mac  	& 85.37 &	 93.75	 &65.62 &	88.28 \\
Rec Pos  	& \underline{71.87}&	     \underline{90.48}	 &\underline{33.33} &	\underline{82.45}\\
\hline
\multicolumn{5}{c}{Deep SNP (Mid+final attention)}\\
\hline
F1 Mac	 	&90.58	 &95.01&	 74.87	& 90.82    \\
F1 Pos 		&83.64	 &92.13&	 53.33	& 86.49	\\		
Prec Mac 	&95.71	 &96.67&	 93.33	& 92.86    \\
Prec Pos 	&\underline{\textbf{100.0}}	 &\underline{\textbf{87.23}}&	 \underline{66.67}	& \underline{\textbf{88.89}}	\\		
Rec Mac  	&85.94    &97.02&	 71.18	& 90.15    \\
Rec Pos  	&\underline{71.87}    &\underline{97.62}&  \underline{44.44}	& \underline{84.22}     \\
\bottomrule
\end{tabular}
\end{sc}
\end{small}
\end{center}
\vskip -0.1in
\end{table}

\subsection{Extended Empirical Results}
\label{subsec:extended_empirical_results}
In this section we visualize the output of the localization unit in Deep SNP.
Please note that these results are produced with the exact same model as used for evaluation and its results are provided in Table~\ref{tab:empirical_results} under \emph{Mid+Final Attention}. Only instead of the final output predictions, we visualized the output of the localization layer.
We need to further clarify that the results of the Section~\ref{subsec:extended_empirical_results} are preliminary and need to be further systematically evaluated and compared to the state-of-the-art.

\begin{figure}[!ht]
\vskip 0.2in
\begin{center}
\centerline{\includegraphics[width=\columnwidth]{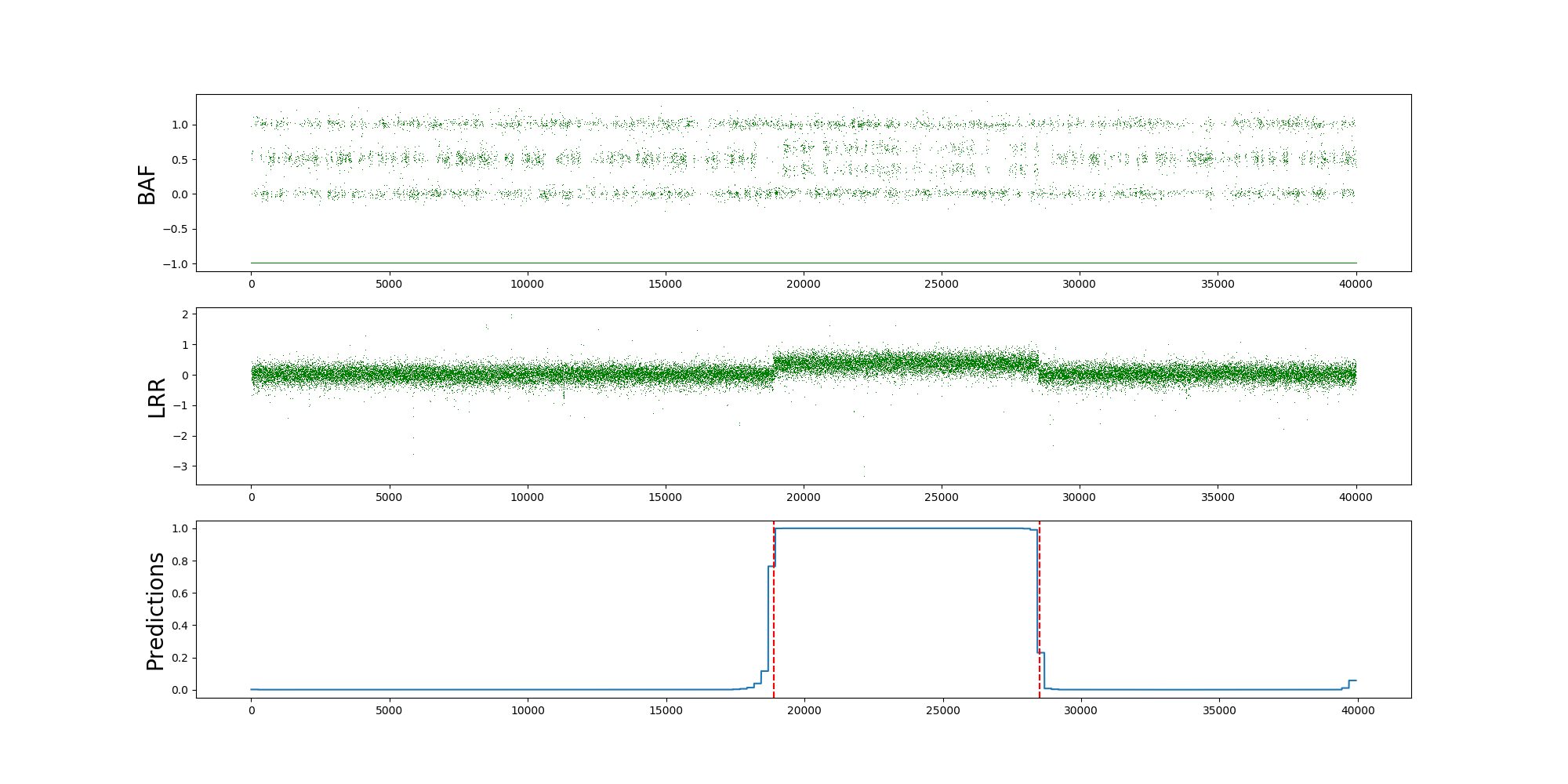}}
\centerline{\includegraphics[width=\columnwidth]{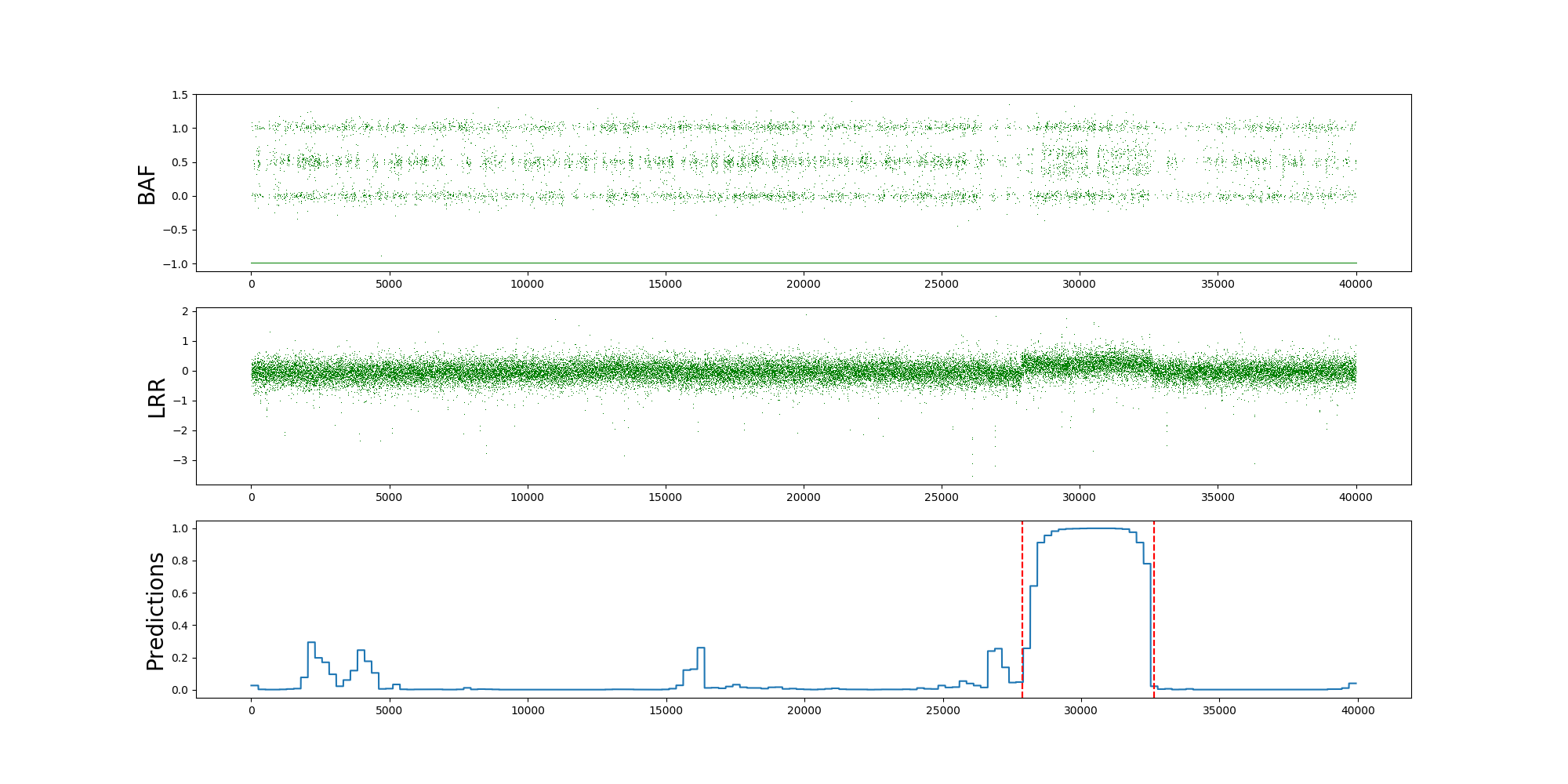}}
\centerline{\includegraphics[width=\columnwidth]{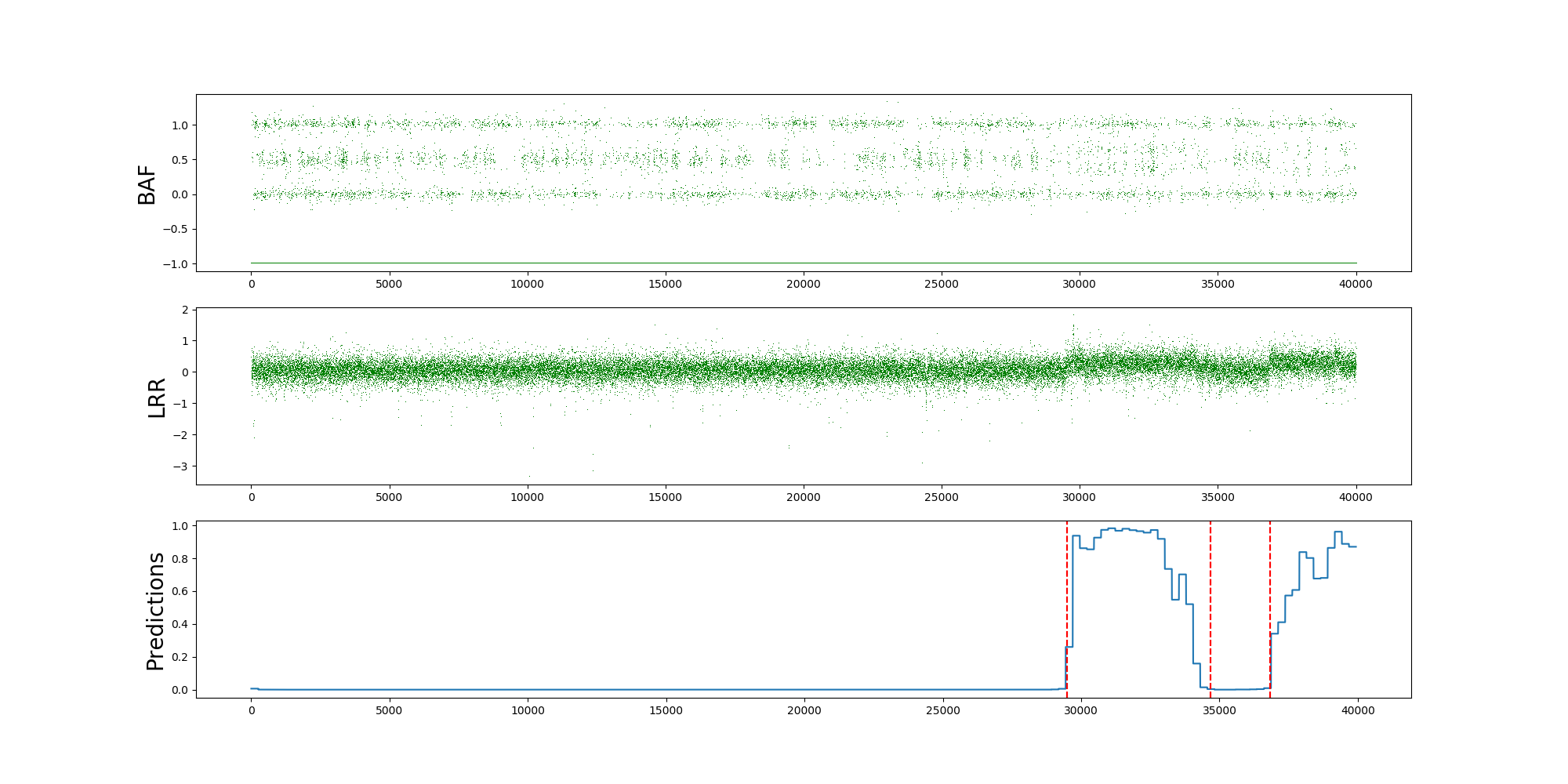}}
\caption{Predictions of the localization unit (with  Mid+Final attention mechanism ) in Deep SNP.  All the 3 windows contain breakpoints. Our model was trained only with whether or not a break point exists in a 40k window. The first row shows the BAF values, the second the LRR values and the third the Deep SNP breakpoint predictions for a given window. The red vertical line represents the truth-set for breakpoints.
}

\label{fig:localisation_preds}
\end{center}
\vskip -0.2in
\end{figure}

\subsection{Discussion}
\label{subsec:discussion}

As can be seen in Table~\ref{tab:empirical_results}, \deepsnp{} outperforms \vgg{}, \densenet{} and \dildensenet{} in terms of F1 and Recall.
The high precision of \vgg{} and \densenet{} are not meaningful as their recall is very low (~15\% and ~4\% for \vgg{} and \densenet{}, respectively) which means 
only a very small portion of breakpoints were detected by them.
\lstmdensenet{} achieves a better recall (on average 43.13\%) but is not still competitive compared to \rc{} and \deepsnp{}.
In addition we observed that as reported in~\cite{gupta_dilated_2017}, incremental dilated convolution layers can be competitive to LSTMs, 
while achieving lower precision. On average, \dildensenet{} achieves 43.32\% average recall while having average precision of 70.54\% compared to average recall of 43.13\% and average precision of 83.87\%.

In contrast, \deepsnp{} manages to achieve high average precisions (82.58\%, 80.75\%, 85.70\% for No Att., Fin. Att and Mid.+Fin. Att, respectively) while having reasonable average recalls (65.9\%, 69.53\%, 74.54\% for No Att., Fin. Att and Mid.+Fin. Att, respectively).
These results suggests that a right combination of dilated convolution with recurrent layers can achieve significantly better performances as used in \deepsnp{}.

\rc{} manages to achieve higher recalls compared to Deep SNP as well as the deep learning baselines. While \rc{} recall is higher than Deep SNP recall, it achieves lower precisions, 
mainly because it detects a substantial higher number of false positive segments (the false positive segments are in reality many, and very short).

As can be seen, the results of Fold 3 in all deep learning models are noticeably lower than the other 3 folds. 
This can be explained by the small training set we used in the current Deep SNP evaluation (9 samples).
Interestingly, \lstmdensenet{} achieves better results on Fold 3 compared to \deepsnp{} and other baselines, despite its poor performance on other Folds compared to \deepsnp{}.
In contrast, \rc{} manages to achieve consistent performances in all the folds, which can be explained by the large amount of samples used compared to other methods. 

These results suggest the possibility that breakpoints available in the Fold 3 might have been different than the other 3 folds (possibly caused by different copy number state transitions as in the other folds).
The performance difference between \rc{} and \deepsnp{} can be also explained by comparing the amount of data used in \rc{} (2.875 samples) and \deepsnp{} (9 samples).
Hence these results could be further improved by providing more training data for \deepsnp{}.

As can be seen in Figure~\ref{fig:localisation_preds} although Deep SNP was only trained with binary weak labels of breakpoints for the whole windows of 40k probes, 
its localization unit is not only capable of accurately pinpointing the correct position of a breakpoint, but also the localization predictions stay at a high value during the whole segment of increased/decreased CNS.

\section{Conclusion}
\label{sec:conclusion}

In this paper, we proposed Deep SNP, a novel Deep Neural Network trained in an end-to-end fashion on SNPa data capable of classifying the presence or absence of one or multiple breakpoints within large genomic windows.

We demonstrated the capabilities of Deep SNP by comparing with state-of-the-art architectures as well as a known biological data analysis tool, \rc{}.
Our results showed Deep SNP outperformed other deep models and could achieve performances that are in a reasonable range compared to specialized tools such as \rc{}.
In addition, we showed qualitative examples from the localization unit that can learn to pinpoint the breakpoint positions as well as the genomic segments although these information were not available to the network during training. Therefore, these promising results are a motivation for our future work to investigate the value of DeepSNP for localizing genomic breakpoints and subsequent or simultaneous prediction of genomic segments.

\section{Future work}
\label{sec:future_work}
In our future work, we will evaluate the results of the localization unit and will compare it with state-of-the-art methods.
Also, we will focus our efforts on improving the recall and also will continue to increase the training data.
In addition, we will investigate the possibilities of the use of localization units for segmentation based on copy number state transition.
Finally, as we believe Deep SNP is capable of learning from other kinds of genomic data, we will apply Deep SNP on other kinds of genomic data such as WGS data.

\section*{Acknowledgements}
This work was carried out within the Austrian Research Promotion Agency (FFG) COIN "Networks" project VISIOMICS and additionally supported by the Austrian Ministry for Transport, Innovation and Technology, the Federal Ministry of Science, Research and Economy, the St. Anna Kinderkrebsforschung and the Province of Upper Austria in the frame of the COMET center SCCH. The authors would like to gratefully acknowledge the support of NVIDIA Corporation with the donation of a Titan X GPU used for this research.

\scriptsize
\bibliography{refs}
\bibliographystyle{icml2018}

\clearpage
\onecolumn
\section{Appendix}

\subsection{Network Architectures}
\label{subsec:architectures}

\begin{table}[!ht]
\begin{center}
\begin{tabular}{cc}
\begin{tabular}{c}
\textbf{BLVGG}\\
Input $2 \times 40k \times 1$ \\
\hline
$1\times10$ Conv(stride-$1\times3$)-$32$-BN-ReLu \\
$1\times5$ Conv(stride-$1\times1$)-$64$-ReLu \\
$1\times5$ Conv(stride-$1\times1$)-$64$-BN-ReLu \\
$1\times2$ Max-Pooling(stride-$1\times2$)+ Drop-Out($0.4$) \\
\hline
$1\times3$ Conv(stride-$1\times1$)-$128$-ReLu \\
$1\times3$ Conv(stride-$1\times1$)-$128$-BN-ReLu \\
$1\times2$ Max-Pooling(stride-$1\times2$)+ Drop-Out($0.3$) \\
\hline
$1\times3$ Conv(stride-$1\times1$)-$256$-ReLu \\
$1\times3$ Conv(stride-$1\times1$)-$256$-ReLu \\
$1\times3$ Conv(stride-$1\times1$)-$256$-BN-ReLu \\
$1\times2$ Max-Pooling(stride-$1\times2$)+ Drop-Out($0.2$) \\
\hline
$1\times3$ Conv(stride-$1\times1$)-$512$-ReLu \\
$1\times3$ Conv(stride-$1\times1$)-$512$-ReLu \\
$1\times3$ Conv(stride-$1\times1$)-$512$-BN-ReLu \\
$1\times2$ Max-Pooling(stride-$1\times2$)+ Drop-Out($0.2$) \\
\hline
$1\times3$ Conv(stride-$1\times1$)-$512$-ReLu \\
$1\times3$ Conv(stride-$1\times1$)-$512$-ReLu \\
$1\times3$ Conv(stride-$1\times1$)-$512$-BN-ReLu \\
$1\times2$ Max-Pooling(stride-$1\times2$)+ Drop-Out($0.2$) \\
\hline
$1\times1$ Conv(stride-$1\times1$)-$2$-ReLu \\
Global-Average-Pooling \\
\hline
$2$-way Soft-Max
\end{tabular}
&
\begin{tabular}{c}
\textbf{BL DenseNet}\\
Input $2 \times 40k \times 1$ \\
\hline
$1\times10$ ConvBn(stride-$1\times5$)-$208$ + DO\\
\hline
Feature Learning Layers (Table~\ref{tab:genet_feature_learning_architecture})\\
\hline
$1\times1$ Conv(stride-$1\times1$)-$2$-ReLu \\
Global-Average-Pooling \\
\hline
$2$-way Soft-Max\\
\hline
\hline
\textbf{BL Dilated DenseNet}\\
Input $2 \times 40k \times 1$ \\
\hline
$1\times10$ DilConvBn(dilation-$1\times3$)-$208$ + DO\\
\hline
Feature Learning Layers with dilation (Table~\ref{tab:genet_feature_learning_architecture})\\
\hline
$1\times1$ Conv(stride-$1\times1$)-$2$-ReLu \\
Global-Average-Pooling \\
\hline
$2$-way Soft-Max\\
\hline
\hline
\textbf{BL LSTM DenseNet}\\
Input $2 \times 40k \times 1$ \\
\hline
$1\times10$ ConvBn(stride-$1\times5$)-$208$ + DO\\
\hline
Feature Learning Layers without dilation(Table~\ref{tab:genet_feature_learning_architecture})\\
\hline
Bi-directional LSTM-$256$\\
\hline
$1\times1$ Conv(stride-$1\times1$)-$2$-ReLu \\
Global-Average-Pooling \\
\hline
$2$-way Soft-Max\\
\end{tabular}
\end{tabular}
\caption{Architecture of the deep learning baselines. BN = Batch Normalization, ReLu = rectified linear unit, Conv = 2d convolution layer. \\ 
Each layer consists of the filter size, layer type, stride size, number of filters and optional layers e.g. $1\times10$ Conv(stride-$1\times3$)-$32$-BN-ReLu describes a 2D convolution layer with filter size = $1\times10$, stride size = $1\times3$, $32$ filters, batch normalization and ReLu activation.}
\label{tab:baselines_architectures}
\end{center}
\end{table}

\begin{table}[!ht]
\begin{center}
\begin{tabular}{c}
Input $2 \times 40k \times 1$ \\
\hline
$1\times10$ DilConvBn(dilation-$1\times5$)-$208$ + DO\\
\hline
Feature Learning Layers (Table~\ref{tab:genet_feature_learning_architecture})\\
\hline
{\begin{tabular}{c|c}Mid-Attention Layer& No-Attention\end{tabular}}
\\ \hline
Bi-directional GRU-$256$
\\ \hline
{\begin{tabular}{c|c}Attention Layer& Softmax\end{tabular}}\\ \hline
Averaging Layer
\end{tabular}
\caption{Architecture of Deep SNP. BN = Batch Normalization, ReLu = rectified linear unit, Conv = 2d convolution layer, ConvBn = Conv+BN+ReLu, DO = Drop-Out($0.2$) \\ 
Each layer consists of the filter size, layer type, stride size, number of filters and optional layers e.g. $4 \times [1\times3$ ConvBn(stride-$1\times1$)-$12$ + DO $]$-$64$ describes a DenseBlock containing $4 \times$ 2D convolution layer with filter size = $1\times3$, stride size = $1\times1$, $12$ filters, batch normalization and ReLu activation, Drop-Out($0.2$) and a final filter size of $64$.}
\label{tab:genet_architecture}
\end{center}
\end{table}

\newcommand{\cross}[1]{#1 $\times$ #1}
\newcommand{\conv}[1]{$\left[\begin{array}{ll} \text{1}\times \text{1} \text{ conv}\\ \text{3}\times \text{3} \text{ conv} \end{array}\right] \times \text{#1}$}
\newcommand{\fcross}[1]{1 $\times$ #1}
\newcommand{\fconv}[1]{$\left[\begin{array}{ll} \bigl[\text{1}\times \text{3} \text{ Conv}\bigr]\text{-}12\\  \text{ concat input} \end{array}\right] \times \text{#1}$}
\begin{table*}[!t]
\centering
\begin{tabular}{c|c}
&{\begin{tabular}{c|c}Deep SNP Feature Learning\textbackslash BL Dilation DenseNet&BLDenseNet\end{tabular}}\\
\hline
Input    &  $2 \times 40k \times1$\\ \hline
Convolution & {\begin{tabular}{c|c}$\bigl[$ \fcross{5} DilConv dil.: \fcross{2}\textbackslash\fcross{9}$\bigr]$-$16$ &$\bigl[$\fcross{5} Conv pad: same stride:\fcross{2}$\bigr]$-$16$\end{tabular}}  \\ \hline
Batch-Normalization    & Batch-Normalization    \\ \hline

\begin{tabular}[c]{@{}c@{}}Dense Block\\ (1)\end{tabular}   & \multicolumn{1}{c}{\fconv{4}}  \\ \hline

\multirow{2}{*}{\begin{tabular}[c]{@{}c@{}}Transition Layer\\ (1)\end{tabular}}   & {\begin{tabular}{c|c}$\bigl[$\fcross{5} DilConv dil.:\fcross{2}, \textbackslash\fcross{27}$\bigr]$-$64$&$\bigl[$\fcross{5} Conv pad: same stride \fcross{1}$\bigr]$-$64$\end{tabular}}  \\ 
\cline{2-2} & \multicolumn{1}{c}{\fcross{2} average pool dil. $1 \times 3$, stride \fcross{2}} \\ 
\cline{2-2}& Batch-Normalization
\\ \hline

\begin{tabular}[c]{@{}c@{}}Dense Block\\ (2)\end{tabular}   & \multicolumn{1}{c}{\fconv{4}}  \\ \hline
\multirow{2}{*}{\begin{tabular}[c]{@{}c@{}}Transition Layer\\ (2)\end{tabular}}   &  {\begin{tabular}{c|c}$\bigl[$\fcross{5} DilConv dil. \fcross{2}, \textbackslash\fcross{81}$\bigr]$-$112$&$\bigl[$\fcross{5} Conv pad: same stride: \fcross{2}$\bigr]$-$112$\end{tabular}}   \\ 
\cline{2-2}   & \multicolumn{1}{c}{\fcross{2} average pool stride \fcross{2}} \\ \cline{2-2}& Batch-Normalization
\\ \hline

\begin{tabular}[c]{@{}c@{}}Dense Block\\ (3)\end{tabular}   & \multicolumn{1}{c}{\fconv{4}}  \\ \hline
\multirow{3}{*}{\begin{tabular}[c]{@{}c@{}}Transition Layer\\ (3)\end{tabular}} &  \multicolumn{1}{c}{$\bigl[$\fcross{2} Conv stride \fcross{2}$\bigr]$-$160$}   \\
\cline{2-2}& \multicolumn{1}{c}{\fcross{2} average pool stride 2} \\ 
\cline{2-2}& Batch-Normalization
\\ \hline

\begin{tabular}[c]{@{}c@{}}Dense Block\\ (4)\end{tabular}   & \multicolumn{1}{c}{\fconv{4}}  \\ \hline
\multirow{3}{*}{\begin{tabular}[c]{@{}c@{}}Transition Layer\\ (4)\end{tabular}} &  \multicolumn{1}{c}{$\bigl[$\fcross{2} Conv, stride \fcross{2}$\bigr]$-$208$}   \\
\cline{2-2}& \multicolumn{1}{c}{\fcross{2} average pool stride 2} \\ 
\cline{2-2}& Batch-Normalization
\\ \hline

\begin{tabular}[c]{@{}c@{}}Dense Block\\ (5)\end{tabular}   & \multicolumn{1}{c}{\fconv{4}}  \\ \hline
\multirow{3}{*}{\begin{tabular}[c]{@{}c@{}}Transition Layer\\ (5)\end{tabular}} &  \multicolumn{1}{c}{$\bigl[$\fcross{2} Conv stride \fcross{2}$\bigr]$-$256$}   \\
\cline{2-2}& \multicolumn{1}{c}{\fcross{2} average pool stride 2} \\ 
\cline{2-2}& Batch-Normalization
\\ \hline

\begin{tabular}[c]{@{}c@{}}Dense Block\\ (6)\end{tabular}  &\multicolumn{1}{c}{\fconv{4}} \\ \hline

\end{tabular}
\vspace{1 ex}
\caption{Feature learning architectures for Deep SNP and BLDenseNet.
}
\label{tab:genet_feature_learning_architecture}
\vspace{-3 ex}
\end{table*}

\end{document}